\documentclass{aa}  

\usepackage{natbib}
\bibpunct{(}{)}{;}{a}{}{,}

\usepackage{graphicx}
\usepackage{txfonts}
\usepackage{hyperref}

\makeatletter
\renewcommand*\aa@pageof{, page \thepage{} of \pageref*{LastPage}}
\makeatother
\hypersetup{colorlinks=true,allcolors=[rgb]{0,0,0.8}}
\usepackage[normalem]{ulem}
\usepackage{color}

\newcommand{\dd}{\mathrm{d}}

\begin{document}

   \title{A non-parametric approach to the relation between the halo mass function and internal dark matter structure of haloes}

   \author{Tamara R. G. Richardson
          \inst{1}
          \and
          Pier-Stefano Corasaniti\inst{1,2}
          }

   \institute{Laboratoire Univers et Théories, Observatoire de Paris, Université PSL, Université de Paris Cité, CNRS, F-92190 Meudon, France\\
              \email{tamara.richardson@obspm.fr}
         \and
             Sorbonne Universit\'e, CNRS, UMR 7095, Institut d'Astrophysique de Paris, 98 bis bd Arago, 75014 Paris, France\\
             }

   \date{Received XXX; accepted YYY}

 
  \abstract
   {Galaxy cluster masses are usually defined as the mass within a spherical region enclosing a given matter overdensity (in units of the critical density). Converting masses from one overdensity definition to another can have several useful applications.}
   {In this article we present a generic non-parametric formalism that allows one to accurately map the halo mass function between two different mass overdensity definitions using 
   the distribution of halo sparsities defined as the ratio of the two masses. We show that changing mass definitions reduces to modelling the distribution of halo sparsities.}
   {Using standard transformation rules of random variates, we derive relations between the halo mass function at different overdensities and
   the distribution of halo sparsities.}
   {We show that these relations reproduce the N-body halo mass functions from the Uchuu simulation  within the statistical errors at a few percent level. Furthermore, these relations allow  the halo mass functions at different overdensities to be related  to parametric descriptions of the halo density profile. In particular, we   discuss the case of the concentration-mass relation of the Navarro-Frenk-White profile. Finally, we   show that the use of such relations allows us to predict the distribution of sparsities of a sample of haloes of a given mass, thus opening the way to inferring cosmological constraints from individual galaxy cluster sparsity measurements.}
   {}

\titlerunning{Relating the mass function and internal structure of haloes}
\authorrunning{T.~R.~G. Richardson \& P.-S. Corasaniti}
   \maketitle
%

\section{Introduction}
\label{sec:intro}
It is now well established that estimates of the abundance of galaxy clusters in the universe can be used to test the standard cosmological scenario \citep{2011ARA&A..49..409A,2012ARA&A..50..353K}. Over the past decade, surveys dedicated to the detection of galaxy clusters have provided complete samples that have enabled numerous cosmological parameter inference analyses using cluster number count measurements \citep{2010ApJ...708..645R,2015MNRAS.446.2205M,2016ApJ...832...95D,2016A&A...594A..24P,2017MNRAS.471.1370S,2018A&A...620A..10P,2019ApJ...878...55B,PhysRevD.102.023509}. In the near future, a new generation of surveys such as  Euclid \citep{Laureijs2011, EuclidI} and the Rubin Observatory's LSST \citep{LSST} will provide larger cluster samples that have the potential to   improve current constraints so as to be complementary to those inferred from other cosmic probes. 

Key to the success of such analyses will be, on the one hand, the ability to control the impact of systematic uncertainties and, on the other hand, the availability of accurate predictions of the halo mass function (HMF) because galaxy clusters are hosted in massive dark matter haloes that are the ultimate result of the hierarchical bottom-up process of cosmic structure formation. Formally, the HMF is the number density of dark matter haloes per unit volume per unit mass, $\dd n/\dd M$, which can be written in the following factorised form \citep[see e.g.][]{Press1974,Bond1991}:
\begin{equation}
    \frac{\dd n}{\dd M}=\frac{\bar{\rho}_m}{M}\frac{\dd \ln\sigma^{-1}}{\dd M}f(\sigma). 
\end{equation}
Here $\bar{\rho}_m$ is the mean cosmic matter density, $\sigma$ is the root-mean-square fluctuation of the linear matter density field smoothed on a spherical region enclosing a mass $M$, and $f(\sigma)$ is the  multiplicity function. The last encodes information on the distribution of halo masses resulting from the non-linear gravitational collapse of matter density fluctuations that leads to the assembly of haloes. However, because of the complexity of this process, predictions of the multiplicity function, and consequently of the HMF, entirely rely on the analysis of cosmological N-body simulations. Numerically calibrated parametrisations of $f(\sigma)$ have been provided in a vast literature  \citep{2001MNRAS.321..372J,2001MNRAS.323....1S,2003MNRAS.346..565R,2006ApJ...646..881W,2007ApJ...671.1160L,Tinker2008,Courtin2011,Angulo2012,Bocquet2016,Despali2016,Diemer2020,Seppi2021}. However, obtaining accurate HMF predictions from numerical simulations poses three main challenges. First of all, simulations must cover large cosmic volumes to resolve with sufficient statistics the high-mass end of the HMF \citep[see e.g.][]{Ishiyama2021}, and possibly to account for the impact of the baryons. In the latter case, this requires the use of N-body or hydrodynamical simulations \citep[e.g.][]{2014MNRAS.440.2290M,2014MNRAS.441.1769C,2014MNRAS.442.2641V,Bocquet2016,2021MNRAS.500.2316C}. Secondly, simulations with different cosmological parameter  set-ups are necessary to evaluate the cosmological dependence (or lack thereof) of the multiplicity function \citep{2001MNRAS.321..372J,Tinker2008,Courtin2011,Despali2016,2019ApJ...872...53M,2019ApJ...884...29N,Diemer2020,2020ApJ...901....5B, Ondaro2021}. Finally, the results depend on the criteria used to detect haloes in the simulations. This is usually done using either the friends-of-friends (FoF) \citep{1985ApJ...292..371D} or spherical overdensity (SO) \citep{1994MNRAS.271..676L} algorithms.
In the first case, haloes are defined as group of particles characterised by an intra-particle distance smaller than a given linking length parameter.  In the second case, haloes correspond to particles within a spherical region that encloses a given overdensity (with respect to the critical or background density). The mass of SO haloes is closer to the  definition of mass that is measured from observations of galaxy clusters. 

In principle, the mass of a galaxy cluster at a given overdensity can be converted to another overdensity if the underlying matter density profile is known. This is the approach originally developed by \citet{Hu2003}, in which the mapping between the mass at two different overdensity values is obtained by assuming the Navarro-Frenk-White \citep[NFW,][]{NFW1997} profile with a given concentration-mass relation. The possibility of mapping halo masses at different overdensities can have several practical applications. As an example, it allows a numerical HMF calibrated for a given overdensity to be transformed into a different one. More specifically, suppose that we have a sample of galaxy clusters with measurements of their spherical mass $M_{500c}$ at an overdensity $\Delta=500\rho_c$, from which we can estimate   the halo mass function, $dn/dM_{500c}$. Suppose that we also have predictions of the HMF for a numerically calibrated multiplicity function $f_{200c}(\sigma)$ using SO halo masses $M_{200c}$ at an overdensity $\Delta=200\rho_c$. Then, we can still make a prediction for $dn/dM_{500c}$ by performing a simple variable transformation:
\begin{equation}
    \frac{\dd n}{\dd M_{500c}}\equiv\frac{\dd n}{\dd M_{200c}}\frac{\dd M_{200c}}{\dd M_{500c}}=\left[\frac{\bar{\rho}_m}{M_{500c}}\frac{\dd \ln{\sigma^{-1}}}{\dd M_{500c}}f_{200c}(\sigma)\right]\frac{M_{500c}}{M_{200c}}.\label{eq:map_example}
\end{equation}
As we can see, this transformation depends crucially on the ratio of the halo masses at two different overdensities. Parametric fits from the analyses of numerical halo catalogues have been provided in the literature for different mass ratios, which have the advantage of being affected by a smaller scatter than the transformation based on the concentration-mass relation \citep[see][]{Bocquet2016, Ragagnin2021}. However, these ratios are not deterministic variables, as implicitly assumed in these studies. Quite the opposite, they are stochastic variables that probe the mass profile of haloes. Dubbed halo sparsities, these ratios were originally investigated in \cite{Balmes2014}, who showed that the ratio of halo masses at two different overdensities provides a proxy of the level of sparsity of the mass distribution inside a halo. Subsequent studies have shown that these ratios encode a considerable amount of cosmological \citep{Corasaniti2018,Corasaniti2021,Corasaniti2022} and astrophysical \citep{Richardson2022} information. 

In this work we present a generic formalism that allows us to accurately map between halo mass functions with different mass overdensity definitions using the distribution of halo sparsities. More specifically, we show that the problem of changing mass definition can be recast into a problem of modelling the distribution of halo sparsities, thus showing the deep connection between the halo mass function at different overdensities and the mass profile of dark matter haloes. This enables us to connect this formalism to the vast literature devoted to the study of the concentration-mass relation of the NFW profile \citep[][]{NFW1997}. Most importantly, we show that such a formalism allows us to accurately predict the distributions of halo sparsities at a given mass using calibrated HMF fitting formula at different overdensities. We demonstrate that this can provide stronger constraints on cosmological parameters than those inferred using average sparsity measurements.

The article is organised as follows. In Section~\ref{sec:simulations} we briefly describe the N-body simulation halo catalogues used as a validation dataset. In Section~\ref{sec:switch_mass_def} we introduce the formalism to map the HMF across different mass overdensity definitions using halo sparsity statistics. We test the accuracy of the formalism against the simulation data and compare to existing results in the literature. In Section~\ref{sec:profile_models} we present the general methodology to convert any model for the internal structure of haloes into a sparsity model and describe the results of specific applications to the NFW  profile. In Section~\ref{sec:cosmo_constraints} we describe a novel method to retrieve cosmological information from a sample of galaxy clusters using measurements of cluster sparsities as function of mass. Finally, in Section~{\ref{sec:conclusion}} we present our conclusions.

\section{Simulation data}
\label{sec:simulations}
We used halo catalogues from the Uchuu suite of N-body simulations \citep{Ishiyama2021}, which were run with the GreeM code \citep{Ishiyama2009, Ishiyama2012}. In particular, we considered haloes with masses $M_{200{\rm c}} > 10^{13}h^{-1}{\rm M}_\odot$ from the large volume $(2h^{-1}{\rm Gpc})^3$ run with $12800^3$ particles (equivalent to a mass resolution of $m_p=3.27\cdot 10^{8}\,h^{-1}\text{M}_{\odot}$), for which the cosmological parameters were set to the values of the \textit{Planck}-CMB 2015 analysis \citep{Planck2015}: $\Omega_m=0.3089$, $\Omega_b=0.0486$, $h=0.6774$, $n_s=0.9667$, and $\sigma_8=0.8159$. 

Halo catalogues were generated with the \textsc{rockstar} code \citep{Behroozi2013a,Behroozi2013b}, which implements a six-dimensional FoF halo finder. The publicly available datasets contain, for each halo in the catalogues, the spherical overdensity halo masses $M_{200{\rm c}}$, $M_{500{\rm c}},$ and $M_{2500{\rm c}}$ at overdensities $\Delta=200,500$, and $2500$, respectively (in units of the critical overdensity). We   used these data to compute the sparsities $s_{200,500}$, $s_{200,2500}$, and $s_{500,2500}$ for each halo in the catalogues. Then, we   estimated the corresponding conditional sparsity distributions $\rho_{\rm s}(s_{\Delta_1, \Delta_2}|M_{\Delta_2})$, the marginal sparsity distributions  $\rho_{\rm s}(s_{\Delta_1, \Delta_2}),$ and the halo mass functions $dn/dM_{\Delta}$. We find the last to be consistent with those presented in the first Uchuu data release \citep{Ishiyama2021}. This dataset is used for all the practical applications of the methods presented hereafter.

\section{Relating sparsity to the halo mass function}\label{sec:switch_mass_def}
In this section we introduce our probabilistic approach to map the HMF from one mass definition to another using halo sparsity. Sparsity is formally defined as \citep{Balmes2014}
\begin{equation}
    s_{\Delta_1,\Delta_2}=\frac{M_{\Delta_1}}{M_{\Delta_2}},\label{eq:sparsity}
\end{equation}
where $M_{\Delta_1}$ and $M_{\Delta_2}$ are spherical masses enclosing overdensities\footnote{As shown in \citet{Balmes2014}, the properties of   halo sparsity are independent of whether overdensities are defined in units of the critical or background density.} $\Delta_1$ and $\Delta_2$, respectively (with $\Delta_2>\Delta_1$). In this context each variable in this expression is treated as a random variable, such that each  can be expressed as the product or ratio of two others:  $M_{\Delta_2}=s_{\Delta_1,\Delta_2}M_{\Delta_1}$ or  $M_{\Delta_1}=M_{\Delta_2}/s_{\Delta_1,\Delta_2}$. Hence, a mapping of the HMF from any of these two mass definitions to the other can be performed using the transformation rules of random variates, which we briefly review in Appendix~\ref{app:random_variables}.

\begin{figure*}
    \centering
    \includegraphics[width = 1\linewidth]{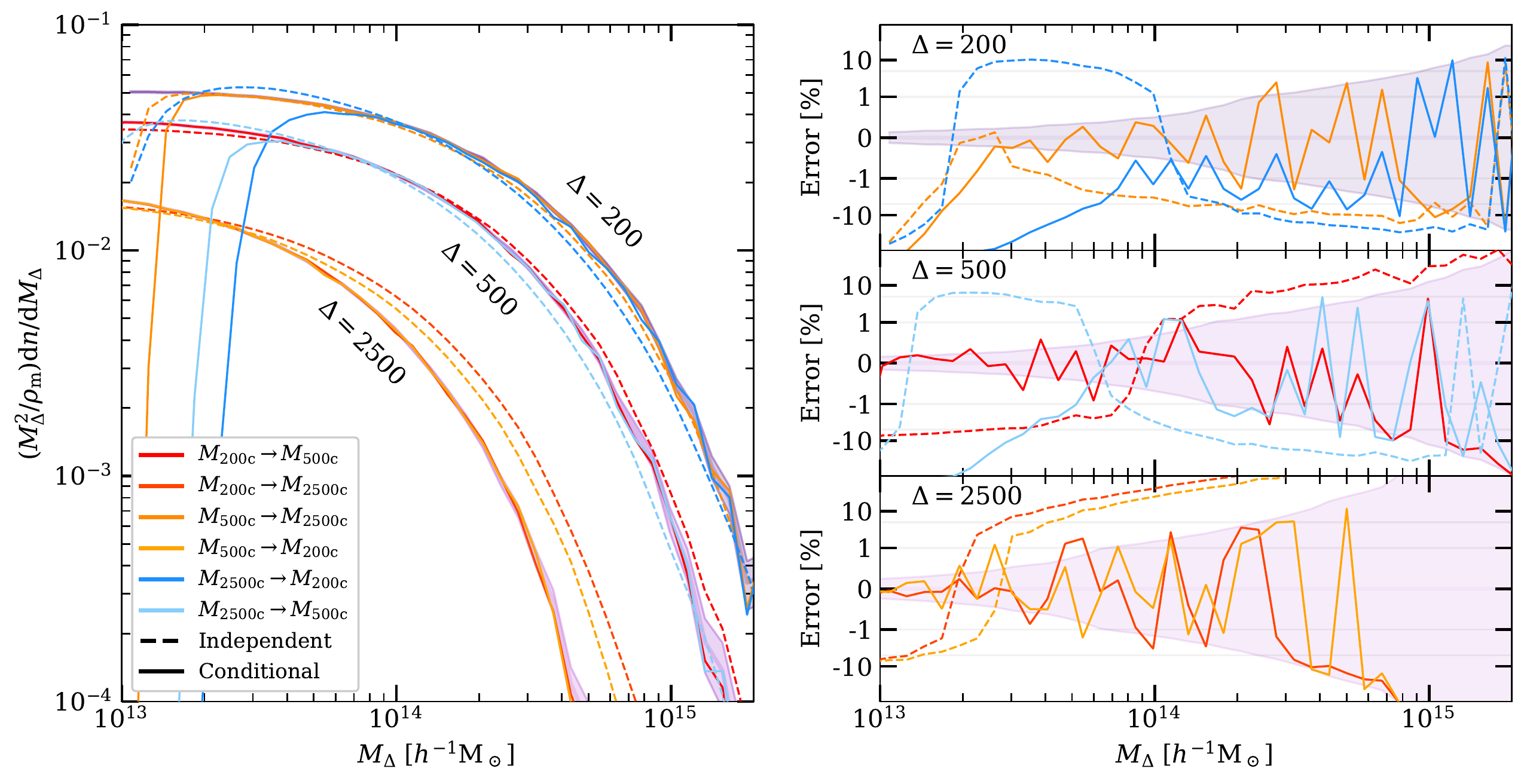}
    \caption{Comparison of the accuracy of the mass dependent HMF transfer formalism with the marginalised formalism. The latter provides poor reconstructions while accounting for the mass dependence results with predictions that are exact to the level of statistical uncertainty. {Left panel:} Estimated HMFs (purple shaded area) at $z=0$ from the Uchuu halo catalogues for  overdensities $\Delta=200$, $500$, and $2500$ (in units of the critical density) plotted against the inward ($200\rightarrow 500$, $200\rightarrow 2500$, and $500\rightarrow 2500$) and outward ($2500\rightarrow 500$, $2500\rightarrow 200$, and $500\rightarrow 200$) reconstructed HMFs from Eq.~(\ref{eq:transfer_down}) and Eq.~(\ref{eq:transfer_up}), respectively, assuming the marginal sparsity distribution (dashed lines) and conditional distribution (solid lines). {Right panels:} Relative error between the reconstructions and the measured HMF at $\Delta=200$ ({top panel}), $\Delta=500$ ({mid panel}), and $\Delta=2500$ ({bottom panel}). The shaded areas around the measured HMFs represents the statistical error on the measurement estimated as the standard deviation over $10^3$ bootstrap iterations. 
    }
    \label{fig:reconstruction}
\end{figure*}

\subsection{Halo mass conversion}
Suppose we want to reconstruct the HMF at the higher overdensity $\Delta_2$ from the HMF at the lower overdensity $\Delta_1$,  which we refer to as inward reconstruction. We let the masses $M_{\Delta_1}$ and $M_{\Delta_2}$ be drawn from $dn/dM_{\Delta_1}$ and $dn/dM_{\Delta_2}$, respectively,  and let the sparsity\footnote{Here we have dropped the indices for ease of reading} ($s$) be drawn from the distribution $\rho_\text{s}(s| M_{\Delta_1})$ conditional to the mass $M_{\Delta_1}$. We note that this distribution is only defined over the interval $1<s < \infty$. Then, as $M_{\Delta_2}$ can be written as the ratio of $M_{\Delta_1}$ to $s_{\Delta_1,\Delta_2}$, the HMF at $\Delta_2$ can be written as a ratio distribution of the two other variables:
\begin{equation}
    \frac{\dd n}{\dd M_{\Delta_2}}(M_{\Delta_2}) = \int_1^{\infty} s\,\rho_\text{s}(s| s M_{\Delta_2})\frac{\dd n}{\dd M_{\Delta_1}}(s M_{\Delta_2})\,\dd s
    \label{eq:transfer_down}
\end{equation}
(see   Eq.~(\ref{eq:ratio_distribution}) in Appendix~\ref{app:random_variables} for a detailed derivation). We note that if we assume that the sparsity distribution is independent of the mass at the outer density $M_{\Delta_1}$, this relation changes only by replacing the conditional distribution of sparsity $\rho_{\rm s}(s|sM_{\Delta_2})$ by its marginal distribution $\rho_{\rm s}(s)$. As such, the only requirement to relate both mass functions is the sparsity distribution.

We now consider the inverse case in which we aim to reconstruct the HMF at the lower density $\Delta_1$ from the HMF at the higher overdensity $\Delta_2$, which we refer to as {outward} reconstruction. Using the definition of the product distribution, see Eq.~(\ref{eq:product_distribution}), we obtain
\begin{equation}
     \frac{\dd n}{\dd M_{\Delta_1}}(M_{\Delta_1}) = \int_{1}^{\infty}\frac{1}{s}\rho_{{\rm s}}(s| M_{\Delta_1}/s)\frac{\dd n}{\dd M_{\Delta_2}}( M_{\Delta_1}/s)\dd s.
     \label{eq:transfer_up}
\end{equation}
Thus, the combination of Eq.~(\ref{eq:transfer_down}) and Eq.~(\ref{eq:transfer_up}) allow us to describe the HMF at a given overdensity contrast as a function of the HMF at any other overdensity.

\subsection{Validation with N-body halo mass functions}\label{HMF_tests}

We test the accuracy of the inward and outward reconstruction given by Eqs.~(\ref{eq:transfer_down}) and (\ref{eq:transfer_up}) against the HMF estimated from the Uchuu halo catalogues at $z=0$ for different overdensity contrasts. 
To do so we numerically estimate the conditional sparsity distributions, $\rho_{\rm s}(s_{\Delta_1, \Delta_2}|M_{\Delta_1})$ and $\rho_{\rm s}(s_{\Delta_1, \Delta_2}|M_{\Delta_2})$, and their marginalised counterparts, from the same halo catalogues used to estimate the HMFs at the two overdensities. These are then used to estimate both sides of Eq.~(\ref{eq:transfer_down}) and Eq.~(\ref{eq:transfer_up}), which we compare
in Fig.~\ref{fig:reconstruction}. In the left panel we plot the N-body mass function at $\Delta=200$ (top curve), $500$ (middle curve), and $2500$ (bottom curve) in units of the critical density against the inward ($M_{200c}\rightarrow M_{500c}$, $M_{200c}\rightarrow M_{2500c}$, and $M_{500c}\rightarrow M_{2500c}$) and outward ($M_{2500c}\rightarrow M_{500c}$, $M_{2500c}\rightarrow M_{200c}$, and $M_{500c}\rightarrow M_{200c}$) reconstructed HMFs assuming the conditional (solid lines) and   marginal (dashed lines) sparsity distributions, respectively. As already mentioned, the latter is equivalent to assuming that the sparsity distribution is independent of the mass at the starting density contrast. In the right panel of Fig.~\ref{fig:reconstruction}, we plot the relative differences with respect to the N-body mass functions at the different overdensities. The shaded areas in both panels correspond to the $1\sigma$ statistical error on the N-body mass functions that we have computed as the standard deviation over $10^3$ bootstrap iterations. 

We can see that using the conditional sparsity distributions nicely reproduces the N-body HMFs within the statistical errors at a few percent level. It is also worth  noting that the inward reconstructions outperform their outward counterparts at the  low-mass end because in the latter case the integration boundaries are below the mass interval over which the HMFs are estimated. This does not occur at the high-mass end due to the presence of the exponential cut-off in the HMFs. In contrast, we find that using the marginal sparsity distribution (i.e. assuming independence), leads to less accurate reconstructed HMFs, which results in relative errors that can exceed the $10\%$ level. In such a case the shape of the recovered HMF more closely resembles that of the one that appears in the integrand of Eq.~(\ref{eq:transfer_down}) or Eq.~(\ref{eq:transfer_up}). Hence, in the case of the inward reconstruction, this results in an underestimation of the reconstructed HMF at low masses and an overestimation at the high-mass end, while the opposite occurs when reconstructing outwards.
In Appendix~\ref{app:validation_higher_z} we present similar tests performed using the Uchuu catalogues at $z=0.5$ and $1$. We find similar trends to those shown in Fig.~\ref{fig:reconstruction}.

\subsection{Validation with analytical results}
\label{subsec:recovering_results}
The general formalism presented above allows us to better understand the relation between halo sparsity and halo mass functions at different overdensities and reproduce past results from the literature. 
As an example, from our formalism we recover a mapping of the form of Eq.~(\ref{eq:map_example}) considered in \citet{Bocquet2016} and \citet{Ragagnin2021}. Such a mapping is equivalent to the inward reconstruction given by Eq.~(\ref{eq:transfer_down}) with the additional assumption that the sparsity distribution is highly peaked about the mean sparsity (i.e. $\langle s_{\Delta_1,\Delta_2}\rangle$). Thus, we approximate the sparsity distribution by a Dirac delta function:
\begin{equation}
    \rho_{\rm s}(s|M_{\Delta_1}) \approx \delta_\text{D}[s - \langle s_{\Delta_1,\Delta_2}\rangle(M_{\Delta_1})].
\end{equation}
Consequently, the integral in Eq.~(\ref{eq:transfer_down}) results in
\begin{equation}
    \frac{\dd n}{\dd M_{\Delta_2}}\approx s_0 \frac{\dd n}{\dd M_{\Delta_1}}(s_0 M_{\Delta_2}),\label{eq:local_balmes}
\end{equation}
where $s_0$ is the root of the argument of the Dirac function (i.e. $s_0-\langle s_{\Delta_1,\Delta_2}\rangle=0$). If the mean sparsity does not vary significantly as function of the halo mass (i.e. $\dd \langle s_{\Delta_1,\Delta_2}\rangle/\dd M_{\Delta_1}\simeq 0$), then $s_0\simeq \langle s_{\Delta_1,\Delta_2}\rangle$. Thus, after some cumbersome algebra, we can write Eq.~(\ref{eq:local_balmes}) as
\begin{equation}
    \frac{\dd n}{\dd M_{\Delta_2}}=\left[\frac{\bar{\rho}_m}{M_{\Delta_2}}\frac{\dd \ln{\sigma^{-1}}}{\dd M_{\Delta_2}}f(\sigma)\right]\frac{1}{\langle s_{\Delta_1,\Delta_2}\rangle},
    \label{eq:leading_order}
\end{equation}
where $\sigma$ is the root-mean-square fluctuation of the linear density field on the mass scale $M_{\Delta_1}=\langle s_{\Delta_1,\Delta_2}\rangle M_{\Delta_2}$. As we can see, for $\Delta_1=200\rho_c$ and $\Delta_2=500\rho_c$ we recover Eq.~(\ref{eq:map_example}). The only fundamental difference is the presence of the expectation value.

Another result we are able to recover is that of the seminal work of \citet{Balmes2014}, which relates the average sparsity to the halo mass functions, thus providing a quantitative set-up to predict the mean sparsity of a cluster sample and to perform cosmological parameter inference analyses \citep[see][]{Corasaniti2018,Corasaniti2021,Corasaniti2022}. Specifically,  given the halo mass function at masses $M_{\Delta_1}$ and $M_{\Delta_2}$, one can infer the value of the average sparsity $s_{\Delta_1,\Delta_2}$ by solving the integral equation
\begin{equation}
    \int \frac{\dd n}{\dd M_{\Delta_2}}\dd \ln{M_{\Delta_2}}=\langle s_{\Delta_1,\Delta_2}\rangle \int \frac{\dd n}{\dd M_{\Delta_1}}\dd \ln{M_{\Delta_1}},\label{eq:balmes14}
\end{equation}
where we have omitted the integration boundaries only for ease of reading. 

We can derive this equation by simply integrating both sides of Eq.~(\ref{eq:transfer_down}) over $\ln{M_{\Delta_2}}$. Then, assuming that sparsity is independent from $M_{\Delta_1}$, we can replace the conditional distribution $\rho_{\rm s}(s|s M_{\Delta_2})$ with the marginal sparsity distribution, $\rho_{\rm s}(s)$ to obtain the following equation:
\begin{equation}
    \int\frac{\dd n}{\dd M_{\Delta_2}}\dd \ln{M_{\Delta_2}}=\int_{1}^{\infty}s\rho_{\rm s}(s)\left[\int \frac{\dd n}{\dd M_{\Delta_1}}(s M_{\Delta_2}) \dd \ln{M_{\Delta_2}}\right]\dd s.
\end{equation}
If the marginal sparsity distribution is peaked around the mean, then we can again approximate, $\rho_{\rm s}(s)=\delta_{\rm D}(s-\langle s_{\Delta_1,\Delta_2}\rangle)$. Finally, by performing the integral over $s$ we recover Eq.~(\ref{eq:balmes14}).

We   conclude this section by emphasising that Eq.~(\ref{eq:local_balmes}) is the mass dependent version of Eq.~(\ref{eq:balmes14}). On the one hand, this shows the deep link between the halo mass function at different overdensities and the halo mass profile. On the other hand, it suggests the possibility of predicting halo sparsity at a given mass from the HMFs. However, rather than using Eq.~(\ref{eq:local_balmes}), this can be done more accurately (as shown by the validation plots of Fig.~\ref{fig:reconstruction}), by assuming an analytical model for the conditional sparsity distribution (e.g. a Gaussian with unknown mean and variance) and solving simultaneously Eq.~(\ref{eq:transfer_down}) and (\ref{eq:transfer_up}) for these two variables as function of halo mass. As we   discuss in Section~\ref{sec:cosmo_constraints}, this allows us to predict the likelihood of individual cluster sparsities that can potentially provide constraints on the cosmological parameters stronger than those inferred using average sparsity measurements.

\section{Halo density profiles}
\label{sec:profile_models}
Halo sparsity is a non-parametric proxy of the halo mass profile. As such, it does not make any assumption on the specific shape of the dark matter density profile. On the other hand, parametric profile parameters can be mapped onto sparsities. Using this in conjunction with its relation to the HMFs, one can map any parametric halo density profile to the HMF at different overdensities. In the following, we   investigate this in the specific case of the NFW profile.

\subsection{Sparsities from the Navarro-Frenk-White profile}
Numerical simulation studies have shown that the density profile of dark matter haloes is described well by a two-parameter fitting function called the NFW profile \citep[][]{NFW1997},
\begin{equation}
    \rho_\text{NFW}(r) = \frac{M_{200\text{c}}}{4\pi\left[\ln(1+c) - c/(1+c)\right]}\times\frac{1}{r\left(\frac{r_{200\text{c}}}{c}+r\right)^2},\label{profNFW}
\end{equation}
where $M_{200c}$ is the mass enclosing the overdensity $\Delta=200$ (in units of the critical density) and $c=r_{200c}/r_s$ is the concentration parameter, which is the ratio of the radius of the spherical region enclosing the mass $M_{200c}$ to the scale radius $r_s$ at which the radial slope of the NFW profile changes from $\propto r^{-1}$ ($r\lesssim r_s$) to $\propto r^{-3}$ ($r\gtrsim r_s$). The concentration parameter provides a simplified description of the radial distribution of mass within haloes since all the information related to a halo's mass assembly history is compressed into a single stochastic variate. It has been the subject of numerous studies in the literature that have investigated its dependence on halo mass, redshift, and cosmology \citep{Bullock2001,2002ApJ...568...52W,2003ApJ...597L...9Z,2004A&A...416..853D,2007MNRAS.378...55M,Zhao2009,Prada2012,Diemer2015,Ludlow2016,Diemer2019,Ishiyama2021,Lopez2022} and its relation to the halo assembly history \citep[see e.g.][]{2003MNRAS.339...12Z,2007MNRAS.379..689L,2007MNRAS.381.1450N,Zhao2009,2012MNRAS.422..185G,2012MNRAS.427.1322L,Ludlow2016,2020MNRAS.498.4450W}. 

Integrating Eq.~(\ref{profNFW}) for a given mass $M_{200c}$ and concentration parameter $c$, one can compute the halo mass at any overdensity $\Delta$, and thus compute the corresponding sparsity from the mass ratio. Hence, as shown in \citet{Balmes2014}, there is a one-to-one relation between the concentration parameter of the NFW profile and the halo sparsity $s^{\rm NFW}_{200,\Delta}$. Specifically, this leads to
\begin{equation}
y^3_{\Delta}\frac{\Delta}{200}=\frac{\ln{(1+c\,y_{\Delta})}-\frac{c\,y_{\Delta}}{1+c\,y_{\Delta}}}{\ln{(1+c)}-\frac{c}{1+c}},\label{eq:sparconc}
\end{equation}
where $y_{\Delta}=r_{\Delta}/r_{200\text{c}}$, with $r_{\Delta}$ being the radius of a sphere enclosing an overdensity $\Delta$, in units of the critical density. Then, solving for $y_{\Delta}$,   the corresponding sparsity is given by
\begin{equation}
s^{\rm NFW}_{200,\Delta}=\frac{200}{\Delta}y_{\Delta}^{-3},
\end{equation}
for any value of $\Delta$ and $c$. We note that by solving this relation for two distinct overdensities, one can calculate any sparsity $s^{\rm NFW}_{\Delta_1,\Delta_2}$. Moreover, this particular relation entails the existence of a continuous differentiable function
\begin{equation}
   s^{\rm NFW}_{\Delta_1, \Delta_2} = f_\text{s}(c)
\end{equation}
and its inverse 
\begin{equation}
    c = f_\text{c}(s^{\rm NFW}_{\Delta_1, \Delta_2})
,\end{equation}
as shown in Fig.~7 of \cite{Balmes2014}. Hence, given that the concentration parameter is a random variate drawn from the conditional distribution $\rho_\text{c}(c|M_{\Delta_1})$, we can derive the conditional distribution of the NFW  sparsity:
\begin{equation}
    \rho_\text{s}(s^{\rm NFW}_{\Delta_1, \Delta_2}|M_{\Delta_1}) = \rho_\text{c}(f_\text{c}(s^{\rm NFW}_{\Delta_1, \Delta_2})|M_{\Delta_1})\left|\frac{\dd f_\text{c}}{\dd s}(s^{\rm NFW}_{\Delta_1, \Delta_2})\right|
    \label{eq:c_to_s}
\end{equation}
(see Eq.~(\ref{eq:transform_1d}) in Appendix~\ref{app:random_variables} for the  derivation). 

The distribution of the concentration parameter is usually modelled as a log-normal density function, whose mean is given by the $c-M$ relation, and a width parameter $\approx 0.25$ \citep[see e.g.][]{Bullock2001,2004A&A...416..853D,2007MNRAS.378...55M}. Thus, given a model for the distribution of the NFW concentration, one can compute the corresponding distribution of the NFW sparsity using Eq.~(\ref{eq:c_to_s}). 

\begin{figure}
    \centering
    \includegraphics[width = 0.95\linewidth]{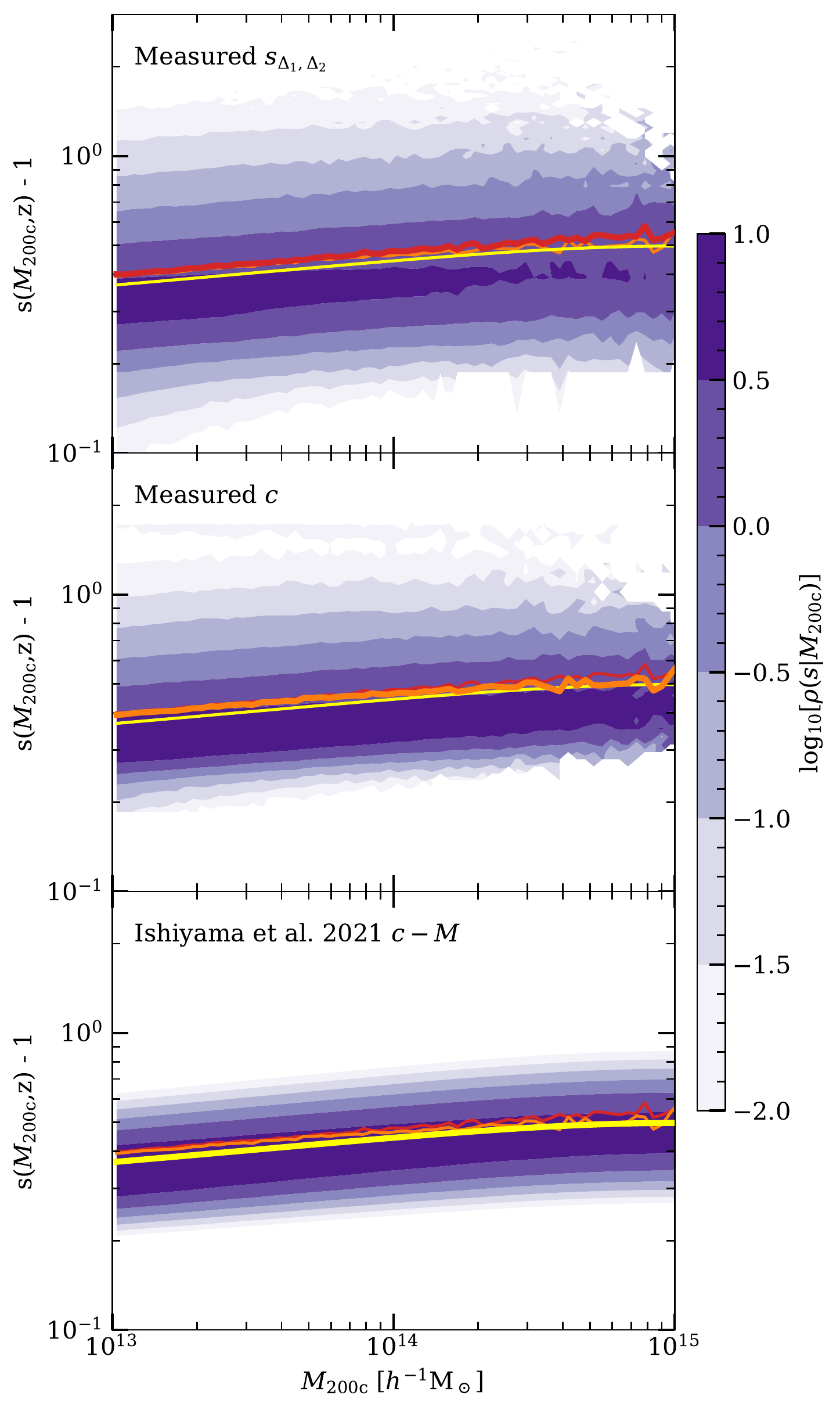}
    \caption{Iso-contours of the conditional density distribution of the halo sparsity, $\rho(s_{200,500}|M_{200 \rm{c}})$. Measurements are from the Uchuu halo catalogue at $z=0$, (top panel), estimated from the distribution of measured concentrations (central panel), and predicted assuming a log-normal distribution of the concentration parameter for which the mean is given by the $c-M$ relation of \citet{Ishiyama2021} calibrated on the Uchuu catalogues and width paramter $\sigma=0.25$ (bottom panel). The coloured lines correspond to the mean of the distribution of measured sparsities (red), and that inferred from the measured concentrations (orange) and from the log-normal distribution (yellow).}
    \label{fig:cMz_conditional}
\end{figure}

In Fig.~\ref{fig:cMz_conditional} we plot iso-contours of the conditional sparsity distribution as function of $M_{200c}$ obtained from the estimated sparsities of the Uchuu halo catalogue at $z=0$ (top panel), the NFW sparsities obtained from the measured concentrations on the same haloes (middle panel), and the NFW sparsities predicted assuming a log-normal concentration distribution for which the mean is given by the concentration-mass relation measured from the analysis of Uchuu haloes \citep{Ishiyama2021} and width parameter $\sigma=0.25$ (bottom panel). The solid lines correspond to the mean values of the distributions: red for the measured sparsities, orange for those inferred from the measured concentrations, and yellow from the log-normal distribution. 

We can see that the last two cases accurately reproduce the mean of the distribution of the sparsities measured from the N-body haloes. However they do not accurately reproduce the scatter around the latter. In particular, we can see that the measured concentrations underestimate the level of scatter for low sparsity values, this is inherently due to assuming a specific shape of the profile, which leads to a loss of information. 
Moreover, further assuming that the concentration follows a log-normal distribution results in a suppression of the scatter on the high-sparsity tail. This is because the log-normal distribution underestimates the distribution of  concentrations at low values, which is primarily sourced by mergers \citep{Richardson2022}.

\begin{figure*}
    \centering
    \includegraphics[width = \linewidth]{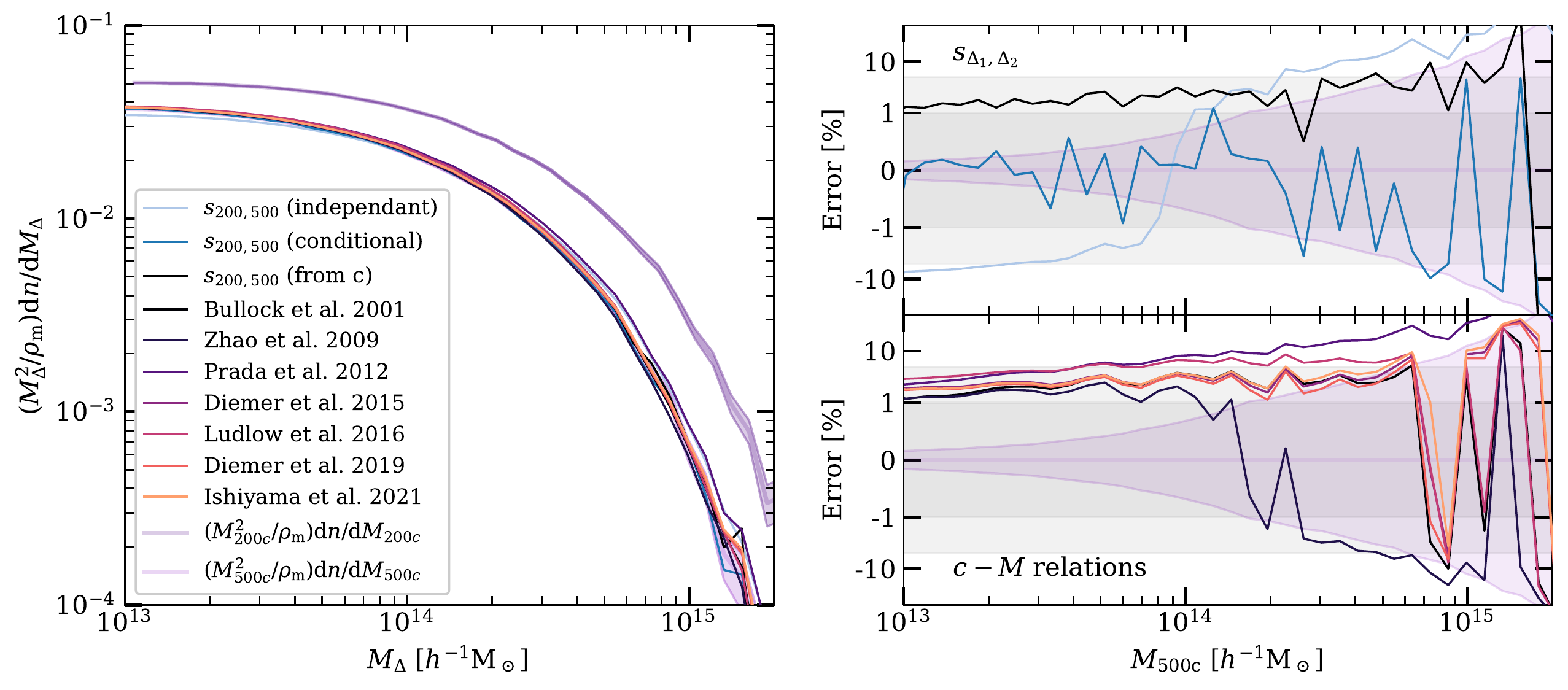}
    \caption{Measuring the effectiveness of transforming the halo mass function from one density contrast to another assuming a $c-M$ relation. {Left panel}:   HMF at $\Delta_1 = 200$ (dark magenta   line) and $\Delta_2 = 500$ (light magenta   line) from the Uchuu halo catalogue at $z=0$ against the reconstructed HMF at $\Delta_2 = 500$ obtained for the different $c-M$ relation models shown in the legend (see text for further information). {Right panel}: Relative error on these reconstructions. As in Fig.~\ref{fig:reconstruction}, the shaded areas around the measured HMFs represent the statistical error on the measurement estimated as the standard deviation over $10^3$ bootstrap iterations.}
    \label{fig:transfer_HMF}
\end{figure*}

\subsection{Halo mass conversions from concentration-mass relations}

Given the relation between halo sparsity and halo concentration, we can map the HMF at different overdensities by combining both the reconstruction procedure presented above in Eq.~(\ref{eq:transfer_down}) and Eq.~(\ref{eq:transfer_up}) with models of the distribution of NFW concentrations from the literature, which are converted into sparsities using using Eq.~(\ref{eq:c_to_s}). This leads to an inward, 
\begin{equation}
    \frac{\dd n}{\dd M_{\Delta_2}}(M_{\Delta_2}) = \int_1^{\infty} s\,\rho_\text{c}(f_\text{c}(s)|s M_{\Delta_2})\left|\frac{\dd f_\text{c}}{\dd s}\right|\frac{\dd n}{\dd M_{\Delta_1}}(s M_{\Delta_2})\,\dd s,
    \label{eq:transfer_down_NFW}
\end{equation}
and outward,
\begin{equation}
     \frac{\dd n}{\dd M_{\Delta_1}}(M_{\Delta_1}) = \int_{1}^{\infty}\frac{1}{s}\rho_\text{c}(f_\text{c}(s)|M_{\Delta_1}/s)\left|\frac{\dd f_\text{c}}{\dd s}\right|\frac{\dd n}{\dd M_{\Delta_2}}( M_{\Delta_1}/s)\dd s,
     \label{eq:transfer_up_NFW}
\end{equation}
reconstruction of the HMF assuming the NFW profile and a $c-M$ relation model. In a similar fashion to how we derived Eq.~(\ref{eq:leading_order}), assuming that the concentration distribution is highly peaked around the $c-M$ relation 
\begin{equation}
    \rho_{\rm c}(c|M_{\Delta_1}) \simeq \delta_D\left[c - \bar{c}(M_{\Delta_1},z)\right],
\end{equation}
one can show that the leading order contributions to Eq.~(\ref{eq:transfer_down_NFW}) and Eq.~(\ref{eq:transfer_up_NFW}) reduce to the formulation of \citet{Hu2003}. Thus, it is clear that the formulation presented above generalises widely used results by including the stochastic natures of the parameters at play, which allows  the study of a wider variety of models within a unified framework.

In Fig.~\ref{fig:transfer_HMF} we perform an inward reconstruction of the HMF at $\Delta_2=500$ starting from the HMF estimated from the Uchuu catalogue at $\Delta_1=200$ assuming 1) the marginal sparsity distribution; 2) the conditional sparsity distribution; 3) the conditional sparsity distribution computed from measured concentrations; and  4) the sparsity distribution predicted assuming a log-normal distribution of the concentration with $\sigma=0.25$ and the mean specified by different $c-M$ relations from \citet{Bullock2001,Zhao2009, Prada2012,Diemer2015,Ludlow2016,Diemer2019,Ishiyama2021}. In addition, we plot the mass functions estimated from the Uchuu halo catalogues
against the reconstructed ones at $M_{500c}$ (left panel) and the relative differences (right panel). As in Fig.~\ref{fig:reconstruction}, the shaded areas correspond to the $1\sigma$ statistical error on the N-body mass function estimated as the standard deviation of $10^3$ bootstrap iterations. 
Again, in the case of the sparsity-based reconstructions, we find that using the conditional sparsity distribution results in a reconstructed HMF that is consistent with that estimated from the N-body halo catalogue within statistical uncertainties with deviations at the sub-percent level up to $M_{500c}\approx 10^{14}\,M_{\odot}h^{-1}$. Instead, using the marginal distribution leads to differences that exceed the $10\%$ level. In the   concentration-based reconstructions, we can see that in the case of the $c-M$ relation from \citet{Prada2012} the reconstructed HMF deviates from the N-body HMF by more than $10\%$ for $M_{500c}\gtrsim 10^{14}\,M_{\odot}h^{-1}$, while in the other cases deviations are within the $1-10\%$ level over the entire mass range. This could be the consequence of a number of factors, such as assumptions in the way the halo concentrations are estimated and the level of scatter we  assume  in the reconstruction. 

To estimate the goodness of the reconstruction for each of the models considered, we compute
\begin{equation}
    \chi^2 = \sum_{i=0}^N \frac{1}{\sigma_i^2}\left[\frac{dn_{\rm rec}}{dM_{500c}}\left(M^i_{500c}\right)-\frac{dn_{\rm N-body}}{dM_{500c}}\left(M^i_{500c}\right)\right]^2,
\end{equation}
where the index $i$ runs over the $N$ mass bins at which the HMF is estimated from the Uchuu haloes and $\sigma_i$ is the corresponding statistical error. We   evaluated the goodness-of-fit of the different models at $z=0.00,0.25,0.50,1.00$, and $2.00$; the  results are quoted in Table~\ref{tab:chi2_cMz}. We find that using the conditional sparsity distribution results in an inward reconstruction that performs significantly better than all other cases at all redshifts. We also find that all reconstructions based on the concentration, including those using the empirical distribution of $c$ from the N-body halo catalogues, exhibit a percent level bias on the reconstruction. We conclude that this bias originates from discrepancies between the true profile and the assumed NFW profile of each halo. Furthermore, we note that among the reconstructions based on the use of $c-M$ relations, the model of \citet{Zhao2009} outperforms the others at low redshifts including the case of \citet{Ishiyama2021}, which was obtained from the analysis of the same simulations. 

\begin{table*}
    \centering
    \caption{$\chi^2$ statistics of the reconstructed HMFs at $\Delta_2 = 500$ and $z=0.00,0.25,0.50,1.00$, and $2.00$ for different reconstruction model assumptions.}
    \begin{tabular}{l|ccccc}
    \hline
        Model & $z = 0$ & $z = 0.25$ & $z = 0.5$ & $z = 1$ & $z = 2$\\
        \hline
        $s_{200,500}$ (independent)  &  329.6 & 289.5 & 363.6 & 243.9 & 8.0 \\
        $s_{200,500}$ (conditional)  &  13.5 & 21.1 & 7.4 & 12.2 & 7.0 \\
        $s_{200,500}$ (from $c$)  &  205.1 & 180.6 & 916.0 & 1902.4 & 382.2 \\
        Bullock et al. 2001  &  310.0 & 237.4 & 158.8 & 896.5 & 664.1 \\
        Zhao et al. 2009  &  135.1 & 66.1 & 70.8 & 184.9 & 1710.3 \\
        Prada et al. 2012  &  1857.6 & 1984.6 & 1728.8 & 553.0 & 32.0 \\
        Diemer et al. 2015  &  321.4 & 152.0 & 55.2 & 196.4 & 76.2 \\
        Ludlow et al. 2016  &  1131.1 & 1282.1 & 953.8 & 121.5 & 188.7 \\
        Diemer et al. 2019  &  258.9 & 181.5 & 79.5 & 111.5 & 71.6 \\
        Ishiyama et al. 2021  &  334.0 & 217.4 & 81.8 & 105.9 & 69.2 \\
        \hline
    \end{tabular}
    \label{tab:chi2_cMz}
\end{table*}

\subsection{Concentration-mass relation from halo mass functions}

An interesting byproduct is the ability to predict the concentration-mass relation from the HMFs at two different overdensities. This can be done using the relation between halo sparsity and HMFs, as well as the relation between the conditional sparsity distribution and that of the concentration. More specifically, in the same fashion used to transform the conditional concentration distribution into the conditional sparsity distribution with Eq.~(\ref{eq:c_to_s}), we perform the inverse operation,
\begin{equation}
    \rho_{\rm c}(c|M_{\Delta_1}) = \rho_{\rm s}(f_{\rm s}(c)|M_{\Delta_1})\left|\frac{\dd}{\dd c}f_{\rm s}(c)\right|.
\end{equation}
Analogously to Sect.~\ref{subsec:recovering_results}, by assuming that the distribution of sparsities is peaked around the mean sparsity value we have
\begin{equation}
    \rho_{\rm s}(s|M_{\Delta_1}) \approx \delta_{\rm D}\left[s - \langle s_{\Delta_1, \Delta_2}\rangle(M_{\Delta_1})\right].
\end{equation}
Hence, this results in a conditional distribution of concentrations that is
also peaked around a value given by
\begin{equation}
    \Tilde{c} := f_{\rm c}(\langle s_{\Delta_1, \Delta_2}\rangle).\label{eq:ctilde}
\end{equation}
Furthermore, using Eq.~(\ref{eq:transfer_up}) we derive an {outward} relation between the HMFs and the mean sparsity:\footnote{As opposed to the {inward} relation, Eq.~(\ref{eq:local_balmes}).}
\begin{equation}
    \frac{\dd n}{\dd M_{\Delta_1}} = \frac{1}{\langle s_{\Delta_1,\Delta_2}\rangle}\frac{\dd n}{\dd M_{\Delta_2}}\left(\frac{M_{\Delta_1}}{\langle s_{\Delta_1,\Delta_2}\rangle}\right).
\end{equation}
Henceforth, given a functional form of the HMFs, we can numerically solve the above equation to obtain $\langle s_{\Delta_1,\Delta_2}\rangle(M_{\Delta_1})$; when  substituted in Eq.~(\ref{eq:ctilde}), this allows us to predict the $c-M$ relation from the HMFs.

\begin{figure}
    \centering
    \includegraphics[width = \linewidth]{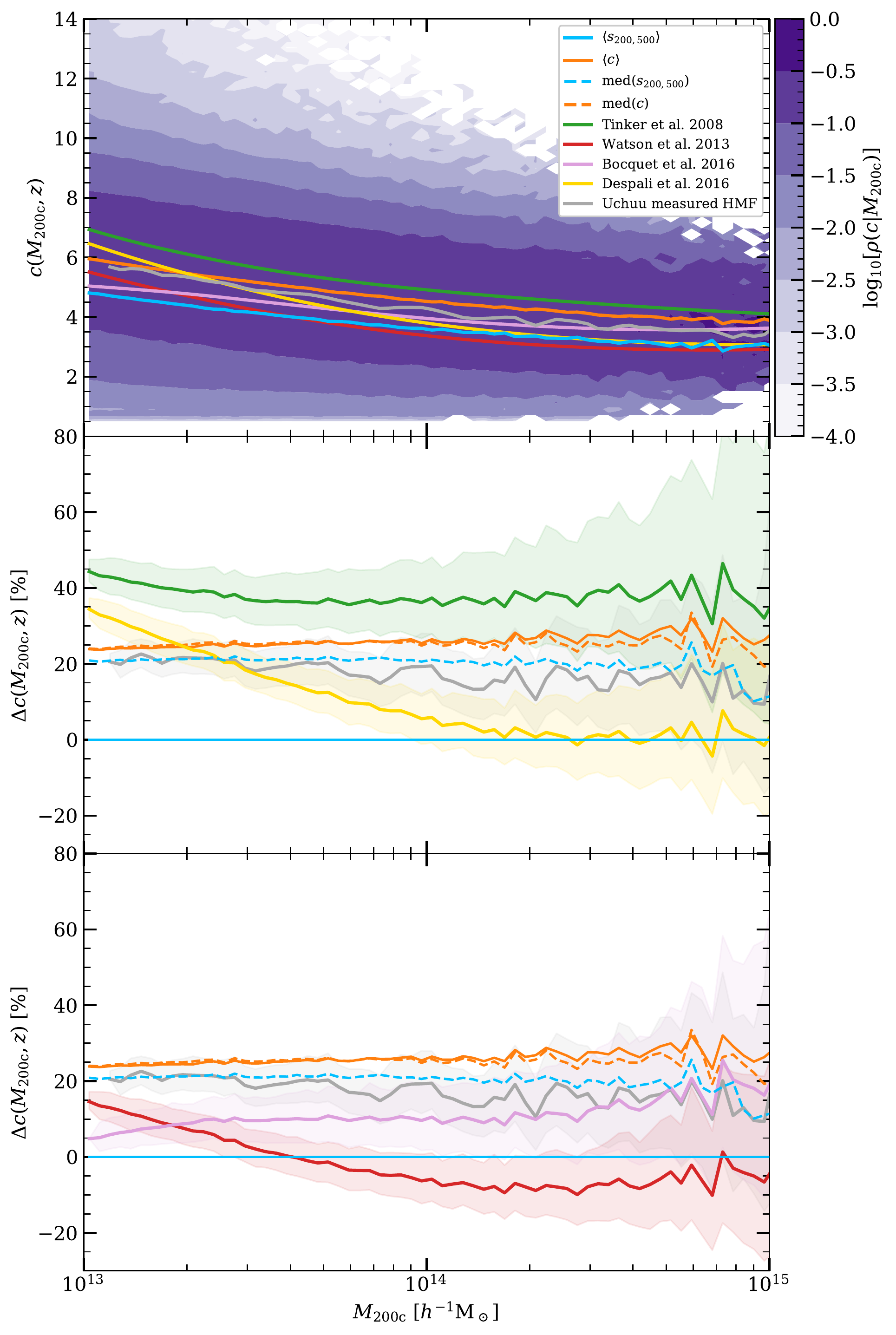}
    \caption{Comparison of the concentration distribution inside the Uchuu simulation, measured $c-M$ relation, and model predictions. {Top panel}: Iso-contours of the conditional concentration $c_{200c}$ from the Uchuu halo catalogues at $z=0$ as a function of $M_{200c}$. The solid lines correspond to the mean $c-M$ relation measured from the concentration (orange line) and mean sparsity $s_{200,500}$ (blue line) of the N-body haloes, and predicted from the measure HMF (grey lines) and HMFs models by \citet{Tinker2008} (green line), \citet{Watson2013} (red line), \citet{Bocquet2016} (pink line), and \citet{Despali2016} (yellow line). {Middle and Bottom panels}: Relative difference between the concentration mass relation predicted from the measured mean sparsity and that measured or predicted using other methods. The shaded area around each model represents one standard deviation around the latter assuming the statistical uncertainty of the HMF measured in the Uchuu simulation. Dashed lines represent the concentration-mass relation predicted from the median sparsity and concentration.
    }
    \label{fig:cM_from_hmf}
\end{figure}

We plot in Fig.~\ref{fig:cM_from_hmf} the mean concentration $c_{200c}$ as a function of $M_{200c}$ from the Uchuu halo catalogue at $z=0$ (solid orange line) with iso-contours of the conditional concentration distribution against the mean $c_{200c}-M_{200c}$ relation obtained from the mean sparsity mass relation $\langle s_{200,500}\rangle(M_{200c})$ measured from the same halo catalogue (solid blue), and that predicted by the HMFs at $\Delta=200\rho_{c}$ and $\Delta=500\rho_{c}$ from 
 \citet{Tinker2008,Watson2013,Bocquet2016,Despali2016} and measured HMFs respectively the green, red, pink, yellow, and grey lines.
 
We find that the predicted mean $c-M$ relations deviates by $10-30\%$ with respect to that estimated from the concentration of the N-body haloes. 
Upon closer inspection we see considerable scatter between the predictions of different HMF prescriptions. This scatter results from the compound effect of model choices, particularly at low masses, and statistical uncertainty on model calibration, especially at high masses, as can be seen in the lower panels of Fig.~\ref{fig:cM_from_hmf}, where we show the relative difference between the $c-M$ relation from measured mean sparsity and those predicted under our assumptions. The shaded areas in this figure correspond to the standard deviation around each model prediction estimated using $10^3$ bootstrap iterations assuming the statistical error on the HMF models is similar to that from the Uchuu simulation.

In addition, we see that the predicted $c-M$ relation from the measured mean sparsity is significantly offset from the prediction from the measured HMF; this is due to our assumption that the distribution is highly peaked around the mean, when it is in fact a wide and highly skewed distribution. We can see this clearly when repeating the same process but using the median instead of the mean. The median, which is closer to the mode of the distribution, is indeed much closer to the prediction from the HMF and is contained within the statistical error around this prediction.
This suggests that, when performing a cosmological parameter inference based upon a prediction of the internal structure of haloes as we do in the following section, the choice of a particular HMF model and the type of statistics considered (i.e. sample mean or median) may introduce systematic errors on the inferred cosmological parameters.

It should be noted that the relations between halo mass functions at different overdensities and the parameters of the parametrised halo density profile discussed here are not limited to the NFW profile, but can be generalised to any parametric profile. For example in Appendix~\ref{app:Einastoprofile}, we discuss the case of the Einasto profile \citep{Einasto1965}.

\section{Forecasting cosmological constraints from individual sparsity measurements}\label{sec:cosmo_constraints}

Cosmological analyses based on cluster sparsity measurements have so far relied on estimates of the ensemble average sparsity of cluster samples at different redshifts \citep[see][]{Corasaniti2018,Corasaniti2021,Corasaniti2022}. However, as shown in Section~\ref{subsec:recovering_results}, by adopting a parametrised form of the conditional sparsity distribution and a parametrisation of the HMFs at two different overdensities, it is possible to predict the mean sparsity and its variance at a given mass and redshift. This provides a quantitative framework to infer cosmological parameter constraints from individual sparsity measurements of galaxy clusters, which may carry more cosmological information than that encoded in the cluster ensemble average, since in the latter case the cosmological signal may be diluted when averaging over the cluster sample. 

We note that while the constraints from sparsity measurements rely on prior theoretical modelling of the HMF, they are to be considered separately from those inferred from number count data analyses. The latter   probes the cosmological imprint encoded in the evolution of the shape and amplitude of the calibrated HMF at the overdensity definition of the cluster observations, while the former tests the differential evolution of the HMF at two overdensities of interest. Formally this is indicated by the presence of the integration variable within Eq.~(\ref{eq:transfer_down}) and Eq.~(\ref{eq:transfer_up}) linking the HMF and sparsity distribution. In essence, the distribution of sparsities controls the difference in the shapes and relative height of the HMFs. Furthermore, studies of the halo concentration \citep[see e.g.][]{2003MNRAS.339...12Z,2007MNRAS.379..689L,2007MNRAS.381.1450N,Zhao2009,2012MNRAS.422..185G,2012MNRAS.427.1322L,Ludlow2016,2020MNRAS.498.4450W} strongly indicate that the internal structures of haloes is linked to their assembly history. Sparsity constraints are thus complementary to number counts, and combining the two probes   provides further improvements to those obtained from sparsity-only analyses. We leave a detailed study of the constraints that can be inferred from the combination of the two probes to future work.

In the following, we assume that the conditional sparsity distribution, $\rho_{\rm s}(s_{\Delta_1,\Delta_2}|M_{\Delta_2})$, is a Gaussian with mean $s_0$ and standard deviation $\sigma_s$. Then, given a parametrised form of the HMFs at redshift $z$ and overdensities $\Delta_1$ and $\Delta_2$, we can simultaneously solve Eqs.~(\ref{eq:transfer_down}) and (\ref{eq:transfer_up}) to derive a prediction for the value of $s_0=\langle s_{\Delta_1,\Delta_2}(M_{\Delta_2},z)\rangle$ and $\sigma_s^2=\sigma^2_{s_{\Delta_1,\Delta_2}}(M_{\Delta_2},z)$. It is worth noting that this particular choice implies that we use the same conditional sparsity distribution for the inward and outward constraints, which in full generality should not be the case. In addition, the distribution of sparsities measured from the N-body halo catalogues appears to be strongly skewed towards high values, and moreover should by definition be $0$ for all values $s_{\Delta_1,\Delta_2} < 1$. While this is far from being verified with our assumptions, the Gaussian distribution is the only distribution that yields a unique solution for this choice of constraints, making it robust to the first guess used to initialise the gradient decent algorithm.

\begin{figure}
    \centering
    \includegraphics[width = \linewidth]{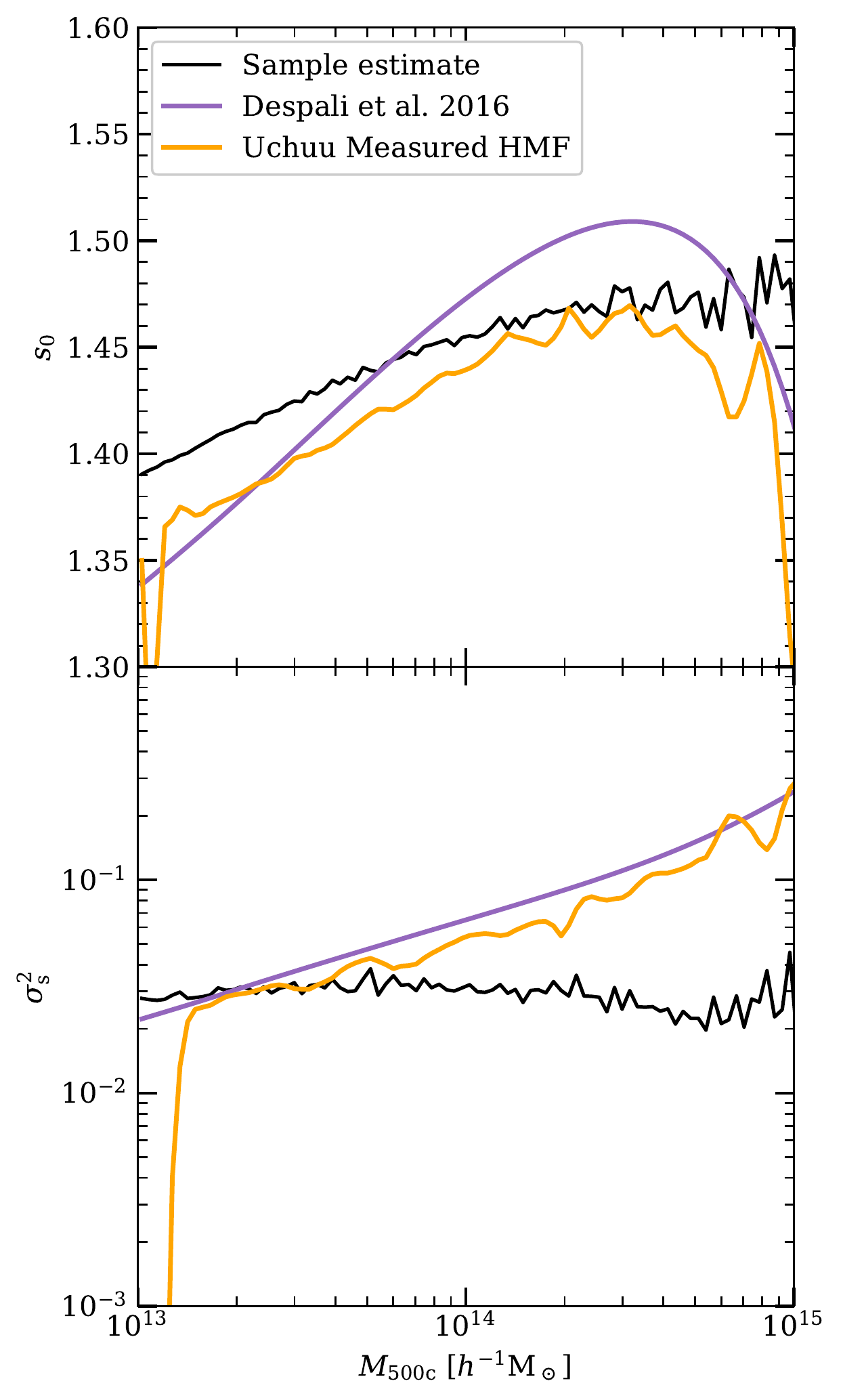}
    \caption{Parameters $s_0$ and $\sigma^2_{\rm s}$ (top and bottom, respectively) for a Gaussian conditional sparsity distribution, $\rho_{\rm s}(s_{200,500}|M_{500 \rm c})$. These parameters are obtained for the distribution that jointly solves the inward, Eq.~(\ref{eq:transfer_down}), and outward, Eq.~(\ref{eq:transfer_up}), reconstructions assuming two HMF models: \cite{Despali2016} (purple) and using the HMF measured from the simulation data (orange). A comparison of  the parameters to the sample mean and variance measured from the data (black lines) shows  that $s_0$ is only accurate to a few percent at recovering the sample mean; this error is carried into the variance, which deviates significantly from the simulation data.}
    \label{fig:mean_var_predict}
\end{figure}

In Fig.~\ref{fig:mean_var_predict} we plot the mean sparsity $s_{200,500}$ and its variance $\sigma_s^2$ in bins of mass $M_{500c}$, as obtained from the analysis of the Uchuu halo catalogue at $z=0$, against the prediction obtained from the HMFs measured from the same sample at $\Delta_1 = 200$ and $\Delta_2 = 500$ in units of the critical density and assuming the analytical fit from \citet{Despali2016}. We can see that $s_0$ is accurate to the order of a few percent when recovering the sample mean. However, we see that the reconstructed variance is significantly biased at high masses. What can be seen is that the variance has only a weak dependency on halo mass while the reconstructed variance increases with mass. This effect is most likely a consequence of the assumptions made on the shape of the probability distribution function since there is no significant difference between using an analytical model for the HMFs and that estimated from the N-body haloes we are trying to reproduce. 

We can now test the level of constraints that can be inferred on the cosmological parameters when using individual sparsity measurements of galaxy clusters. To this end, we generated a synthetic dataset consisting of 118 cluster-scale haloes ($M_{200 \rm{c}} > 10^{14}h^{-1}{\rm M}_\odot$) randomly selected over all Uchuu catalogues up to $z=0.63$. This particular selection was done so as to have a crude resemblance to the CHEX-MATE cluster sample \citep{CHEX-MATE2021}. For each of these haloes we   computed the sparsity $s_{200,500}$. We compared the constraints from the individual sparsity measurements to those from the ensemble average estimates at different redshifts \citep[see e.g.][]{Corasaniti2018}. For this purpose we split the synthetic sample into $N_z = 6$ independent redshift bins and computed the average sparsity in each of them. 

In order to evaluate the differences between the two approaches, we first consider an ideal case in which we neglect uncertainties on the sparsity measurements and assume a Gaussian likelihood function:
\begin{equation}
    \ln\mathcal{L} =  -\frac{1}{2}\sum_{i = 1}^N\left\{\ln\left[2\pi\,\sigma_{\rm s}^2(M_i, z_i)\right] + \frac{\left[s_i - s_0(M_i, z_i)\right]^2}{\sigma_{\rm s}^2(M_i, z_i)}\right\}.
    \label{eq:indep_loglike}
\end{equation}
Here $s_i$ is the sparsity of the $i$-th synthetic data point with $N=118$, $s_0(M_i,z_i)$ and $\sigma_\mathrm{s}^2(M_i,z_i)$ are respectively the mean and variance of sparsities at a given mass and redshift  as predicted for a given set of cosmological parameters by simultaneously solving Eq.~(\ref{eq:transfer_down}) and (\ref{eq:transfer_up}), with HMFs given by the analytical fit of \citet{Despali2016} and assuming, $\rho_{\rm s}(x|M_{i}, z_i)$, to be Gaussian with mean, $s_0(M_i,z_i)$, and variance $\sigma_\mathrm{s}^2(M_i,z_i)$. The cosmology dependence of the likelihood is captured through that of the HMF at the density contrasts of interest. In the case of \citet{Despali2016} this dependence is embodied by the variation of the fit parameters with the virial overdensity contrast. Moreover, this choice is motivated by the need of a HMF definition compatible with matched haloes. In the case of the ensemble average sparsity measurements, the sum in Eq.~(\ref{eq:indep_loglike}) runs over the redshift bins (i.e. $N=6$), and the average sparsities are compared to the theoretical expectation \citep[][]{Corasaniti2018} through a Gaussian likelihood with variance $\sigma^2 = 0.2^2$. We focus on $\Omega_m$ and $\sigma_8$ and use affine invariant Markov chain Monte Carlo sampling \citep{Goodman2010, emcee2013} of the log-likelihood with uniform priors $0.1 < \Omega_m < 0.6$ and $0.3<\sigma_8<1.3$. 

\begin{figure}
    \centering
    \includegraphics[width = \linewidth]{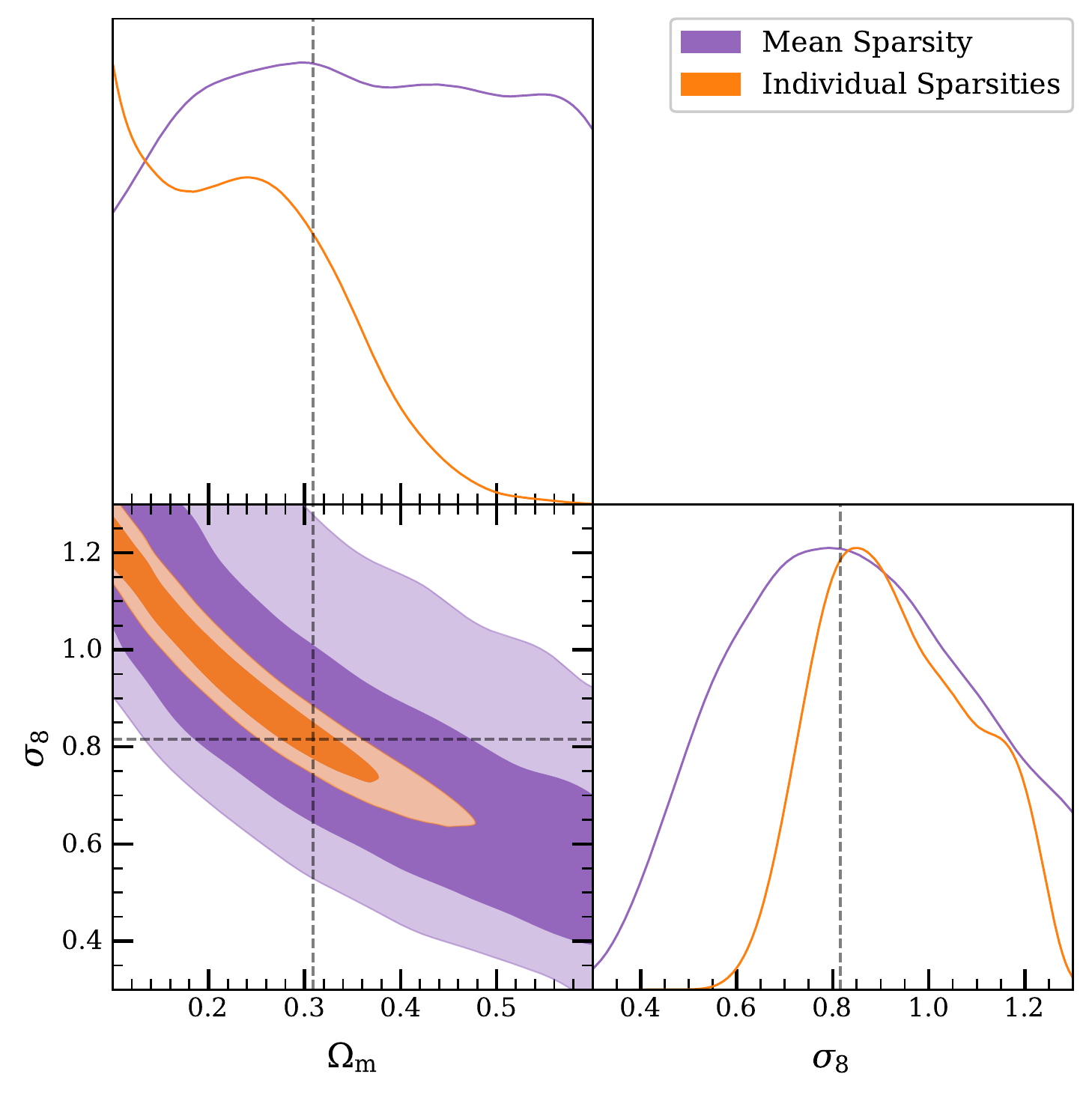}
    \caption{Posterior distributions resulting from the analysis of   118 randomly selected haloes from the Uchuu simulation. Shown in  purple is  the methodology of \cite{Corasaniti2018} that calculates the mean sparsity in $N_z = 6$ redshift bins and in orange the method where the haloes are treated as an individual data point (see  Sect.~\ref{sec:cosmo_constraints}). In both cases the same input information is used (i.e. the same 118 haloes and using the HMF definition of \citealt{Despali2016}). There is  a clear increase in the constraining power when using the second method, this is simply due to avoiding the information loss that occurs when binning and calculating the mean sparsity.}
    \label{fig:posterior}
\end{figure}

\begin{figure}
    \centering
    \includegraphics[width = \linewidth]{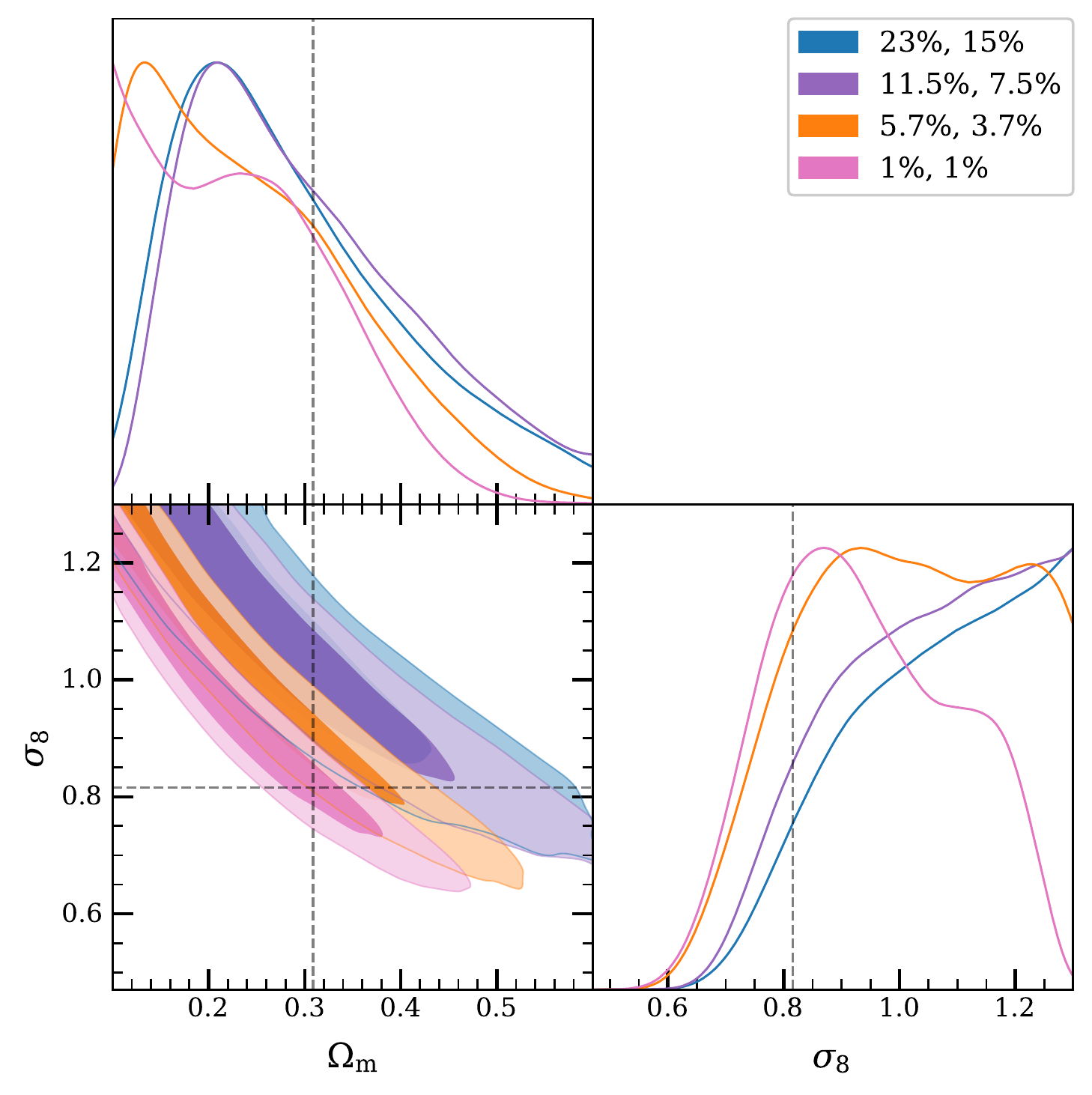}
    \caption{Posterior distributions resulting from the analysis of a sample of 118 randomly selected haloes from the Uchuu simulation modelling measurement errors. Each contour corresponds to a model for the relative errors on clusters masses ($\delta M_{200{\rm c}, i}/M_{200{\rm c}, i},\ \delta M_{500{\rm c}, i}/M_{500{\rm c}, i}$):  in blue (23\%, 15\%) the magnitude of errors estimated for the CHEX-MATE sample; in purple (11.5\%, 7.5\%); in orange (5.7\%, 3.7\%); and in pink (1\%, 1\%). For the smallest errors the posterior from Fig.~\ref{fig:posterior} is recovered where errors were neglected. Also seen is  that a naive modelling of errors induces a bias towards increasingly large values of $S_8$.}
    \label{fig:posterior_errors}
\end{figure}

In Fig.~\ref{fig:posterior} we show the resulting posterior distributions. In both cases we see that the Uchuu simulation's fiducial cosmology is recovered within the 1$\sigma$ contour of each posterior. Moreover, consistently with results from previous studies, the sparsity constraints line up along the $S_8 = \sigma_8 \sqrt{\Omega_m/0.3}$ degeneracy curve. We may also find that using individual sparsity measurements rather than the ensemble averages at different redshifts leads to much stronger constraints. This is due to avoiding the information loss caused by binning and calculating the mean sparsity in each redshift bin. However, this comes at the cost of increased run time resulting from the complexity of the likelihood evaluation. 

In order to account for sparsity measurement errors due to observational uncertainties of the cluster masses, we now assume for simplicity that individual mass measurements are drawn from independent log-normal distributions of mean $M_{\Delta_i}$ and variance $\delta M_{\Delta_i}^2$. From this, we obtain the joint distribution of errors on the sparsity, $s_{200,500}$ and the inner mass, $M_{500 \rm c}$, using a ratio distribution, Eq.~(\ref{eq:ratio_distribution}), over which we marginalise the likelihood function. This simple approach is sufficient when the errors on the masses are small, typically a few percent; however, if the errors are larger, the resulting error distribution assigns significant probabilities to sparsities  $s_{200,500} < 1$, a non-physical regime. This has the systematic effect of assigning weight to low sparsities and greatly biasing the likelihood towards large values of $S_8$. Accurate error modelling, in particular the correlation between the errors, is therefore crucial to avoid this statistical induced bias.

With the intent of diminishing this effect we add the following prior to our error model,
\begin{equation}
    \rho_{\rm p}(s) = 1- \exp\left[-\frac{1}{2}\frac{(s - 1)^2}{s_i^2(\delta M_{200 {\rm c}, i}^2/M_{200 {\rm c}, i}^2 + \delta M_{500 {\rm c}, i}^2/M_{500 {\rm c}, i}^2)}\right],
\end{equation}
which reduces the non-physical weight placed on low sparsities to produce the posteriors of Fig.~\ref{fig:posterior_errors}. We note the distinction between $s_i$ (the measured sparsity) and $s$ (the variable over which we marginalise the error distribution). We chose to adapt the width of this prior with the magnitude of the errors so as to correct the low-mass error regime as little as possible. We produce posterior distributions for four error models. In blue we show the case of cluster mass errors estimated by \cite{Corasaniti2022} for the CHEX-MATE sample, $\delta M_{200{\rm c}, i}/M_{200{\rm c}, i} = 0.23$, $\delta M_{500{\rm c}, i}/M_{500{\rm c}, i} = 0.15$; in purple is shown the case where we halve these errors; for the orange contours we have reduced the original errors by a factor of 4; and  in pink we use percent level errors. What can be clearly seen is the effect of the bias induced by the crude modelling of errors. This bias is naturally reduced when we consider smaller errors on the cluster masses, with the case with the smallest errors recovering the contours obtained in the ideal case (i.e. with no mass measurement uncertainties).

While it is difficult to conclude on the case with the largest errors as the prior strongly influences this specific result, we do note that models with errors comparable to those of  upcoming missions already produce stronger constraints than if we consider only the ensemble average sparsity. It is also worth noting that the simplifying assumptions that produced these forecasts can be alleviated with known methodologies. For example, we could replace the analytical form of the sparsity distribution with one predicted by a cosmological emulator trained over a large sample of cosmological simulations. Moreover, accurately modelling the mass measurement errors can further improve the cosmological constraints providing a new avenue for testing cosmology.

\section{Conclusions}
\label{sec:conclusion}%

 It is currently widely accepted that observations of galaxy clusters provide exceptional opportunities to study both cosmology and astrophysics. While recent cosmological studies using galaxy clusters have been primarily focused on cluster number counts, the internal structure of dark matter haloes, as probed by halo sparsity, has proven to be a new and useful probe for both cosmology \citep{Balmes2014, Corasaniti2018,Corasaniti2021, Corasaniti2022} and the astrophysics of galaxy clusters \citep{Richardson2022}, thanks to current and upcoming observations of galaxy clusters reaching the level of precision required to extract this information encoded in the mass profile of clusters.

In this paper we investigated how sparsity statistics can be further used to map the relation between two halo mass functions estimated at two distinct density contrasts. Within a probabilistic framework we were able to exactly relate both halo mass functions using only the distribution of sparsities conditional to halo mass. In particular, we showed that with additional assumptions on this distribution we were able to recover formulations previously used in the literature. Moreover, we demonstrated that it is also possible to retrieve information about the sparsity distribution directly from the halo mass functions.

The non-parametric nature of halo sparsity also allowed us to express the mapping between halo mass functions in terms of any parameters describing the density profiles of haloes. To this end, we examined the specific case of NFW concentration. Thus, we showed that using the relation between sparsity and concentration it is possible to map the halo mass function to any density contrast simply by assuming a $c-M$ relation, and inversely to predict a $c-M$ relation given the HMF at two overdensity contrasts.

Finally, we showed that our method for predicting the distribution of sparsities at any mass, redshift, and cosmology can be directly applied to perform cosmological inference analyses and provide significantly stronger constraints than current methods based on the use of ensemble average sparsity measurements. However, the method presented here can  be further expanded through the use of emulators and more accurate handling of the cluster mass measurement errors.

This project made use of publicly available data from the Skies and Universes database.\footnote{\href{http://skiesanduniverses.org/Simulations/Uchuu/}{http://skiesanduniverses.org/Simulations/Uchuu/}} In addition, many of the Python codes and transformed data products used throughout this project are made publicly available online.\footnote{\href{https://gitlab.obspm.fr/trichardson/HMF-relations-and-sparsity-predictions}{https://gitlab.obspm.fr/trichardson/HMF-relations-and-sparsity-predictions}}

\begin{acknowledgements}
We thank Yann Rasera, Amandine Le Brun and the anonymous referee for their insightful comments on this manuscript. This work has made use of the Infinity Cluster hosted by Institut d'Astrophysique de Paris. We thank Stephane Rouberol for running smoothly this cluster for us.

We thank Instituto de Astrofisica de Andalucia (IAA-CSIC), Centro de Supercomputacion de Galicia (CESGA) and the Spanish academic and research network (RedIRIS) in Spain for hosting Uchuu DR1 and DR2 in the Skies \& Universes site for cosmological simulations. The Uchuu simulations were carried out on Aterui II supercomputer at Center for Computational Astrophysics, CfCA, of National Astronomical Observatory of Japan, and the K computer at the RIKEN Advanced Institute for Computational Science. The Uchuu DR1 and DR2 effort has made use of the skun@IAA\_RedIRIS and skun6@IAA computer facilities managed by the IAA-CSIC in Spain (MICINN EU-Feder grant EQC2018-004366-P).

We thank the developers and maintainers of the \texttt{colossus} package \cite{Colossus} that was used in this work.
\end{acknowledgements}

\bibliographystyle{aa}
\bibliography{bibliography}

\begin{appendix} 
\section{Transformation of random variates}
\label{app:random_variables}
Throughout this work we treat halo properties as random variables. As such, each variable is associated with a probability distribution function (PDF). When we apply a transformation to the random variable, the PDF  must also be transformed. 

We let $X$ and $Y$ be two random variates drawn respectively from $\rho_x(x)$ and $\rho_y(y)$ and related through a deterministic function, $Y = f(X)$. Due to the conservation of probability, $\rho_y(y)\dd y = \rho_x(x)\dd x$, we can relate the two  PDFs,
\begin{equation}
    \rho_y(y) = \rho_x(f^{-1}(y))\left|\frac{\dd f^{-1}}{\dd y}\right|,
    \label{eq:transform_1d}
\end{equation}
assuming the transformation to be invertible.

Within the context of this work we are interested in transformations involving two random variates:  $Z = f(X,Y)$. Relating the PDF of $Z$ to the joint distribution, $\rho_{xy}(x,y)$ of $X$ and $Y$, requires additional thought compared to the one-dimensional case.  In most cases the function $f(X,Y)$ will not be invertible. However, this can  be circumvented through the introduction of a fourth variable $W$. We define two column vectors,
\begin{equation}
\left[\begin{matrix}
Z \\
W
\end{matrix}\right] = \vec{f}(X,Y) =
\left[\begin{matrix}
f_Z(X,Y) \\
f_W(X,Y)
\end{matrix}\right]
\end{equation}
and
\begin{equation}
\left[\begin{matrix}
X \\
Y
\end{matrix}\right] = \vec{g}(Z,W) =
\left[\begin{matrix}
g_X(Z,W) \\
g_Y(Z,W)
\end{matrix}\right]
,\end{equation}
as the transformations between these variables. Through the conservation of probability, the joint distribution, $\rho_{zw}(Z,W)$, can be written as
\begin{equation}
    \rho_{zw}(z,w) = \rho_{xy}[g_X(z,w), g_Y(z,w)]\left|
    \begin{matrix}
    \partial_z g_X & \partial_w g_X\\
    \partial_z g_Y & \partial_w g_Y
    \end{matrix}\right|.
    \label{eq:transform_2d}
\end{equation}
The distribution for $Z$ can then be obtained by marginalising over $W$:
\begin{equation}
    \rho_z(z) = \int \rho_{zw}(z,w) \dd w.
\end{equation}

In this work we are particularly interested in the PDF of the product, $Z =XY$, and ratio, $Z = \frac{X}{Y}$, of two random variables. In the case of the product, we define
\begin{equation}
\left[\begin{matrix}
Z \\
W
\end{matrix}\right] = 
\left[\begin{matrix}
XY \\
Y
\end{matrix}\right] \;\text{and}\;
\left[\begin{matrix}
X \\
Y
\end{matrix}\right] = 
\left[\begin{matrix}
Z/W \\
W
\end{matrix}\right] 
\end{equation}
as the transformation between the four random variables. We can then write
\begin{equation}
    \rho_z(z) = \int\frac{1}{|w|}\rho_{xy}(z/w, w)\dd w
    \label{eq:product_distribution}
,\end{equation}
the PDF of $Z$. The ratio $Z = X/Y$ similarly leads to
\begin{equation}
\left[\begin{matrix}
Z \\
W
\end{matrix}\right] = 
\left[\begin{matrix}
X/Y \\
Y
\end{matrix}\right] \;\text{and}\;
\left[\begin{matrix}
X \\
Y
\end{matrix}\right] = 
\left[\begin{matrix}
ZW \\
W
\end{matrix}\right] .
\end{equation}
This results in the ratio distribution
\begin{equation}
    \rho_z(z) = \int|w|\rho_{xy}(zw, w)\dd w.
    \label{eq:ratio_distribution}
\end{equation}

\section{Validation against N-body halo catalogues at \texorpdfstring{$z>0$}{z>0}}
\label{app:validation_higher_z}
\begin{figure*}
    \centering
    \includegraphics[width = 1\linewidth]{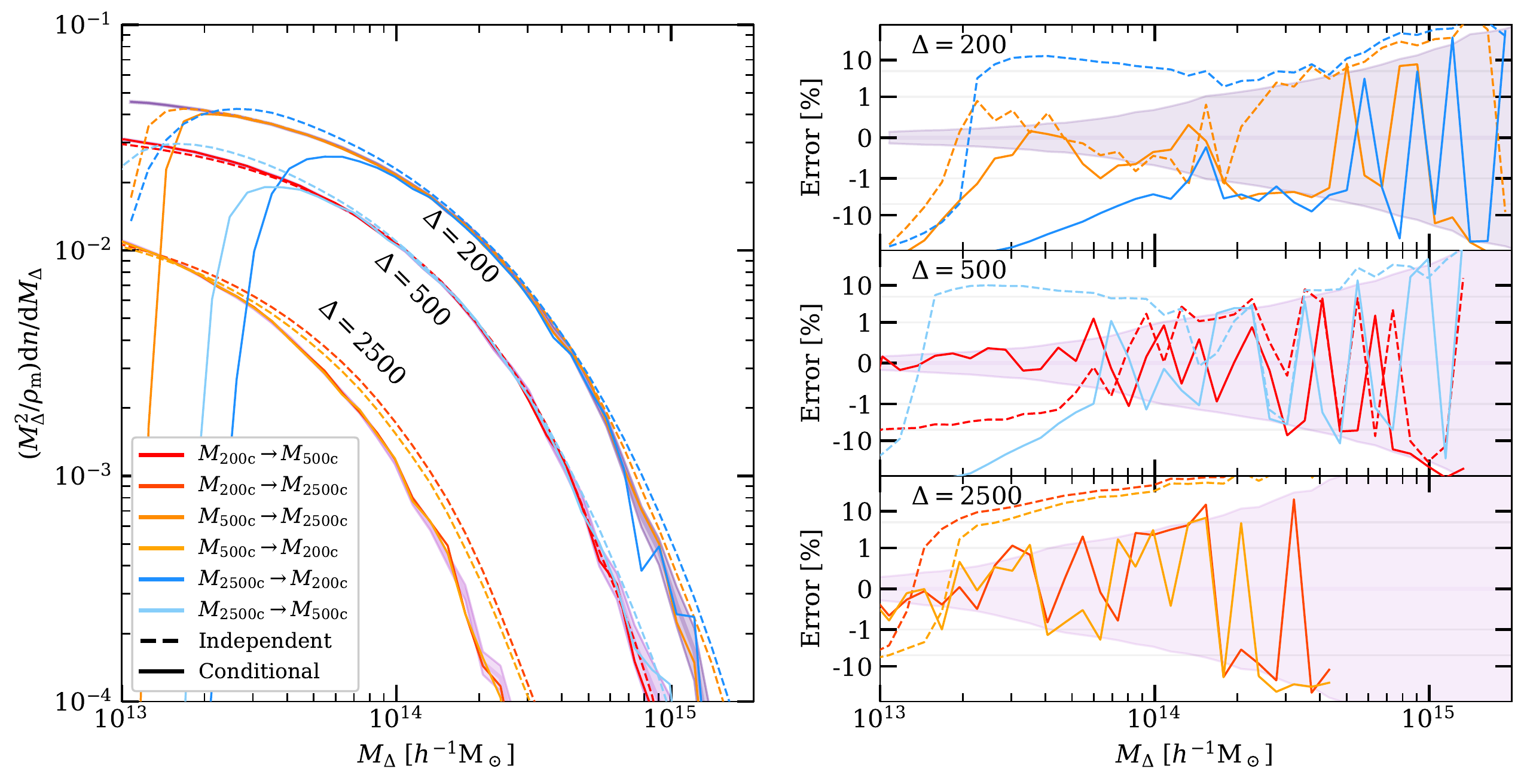}
    \caption{Same as Fig.~\ref{fig:reconstruction}, but at redshift $z= 0.5$. 
    }
    \label{fig:reconstruction_z0p49}
\end{figure*}

\begin{figure*}
    \centering
    \includegraphics[width = 1\linewidth]{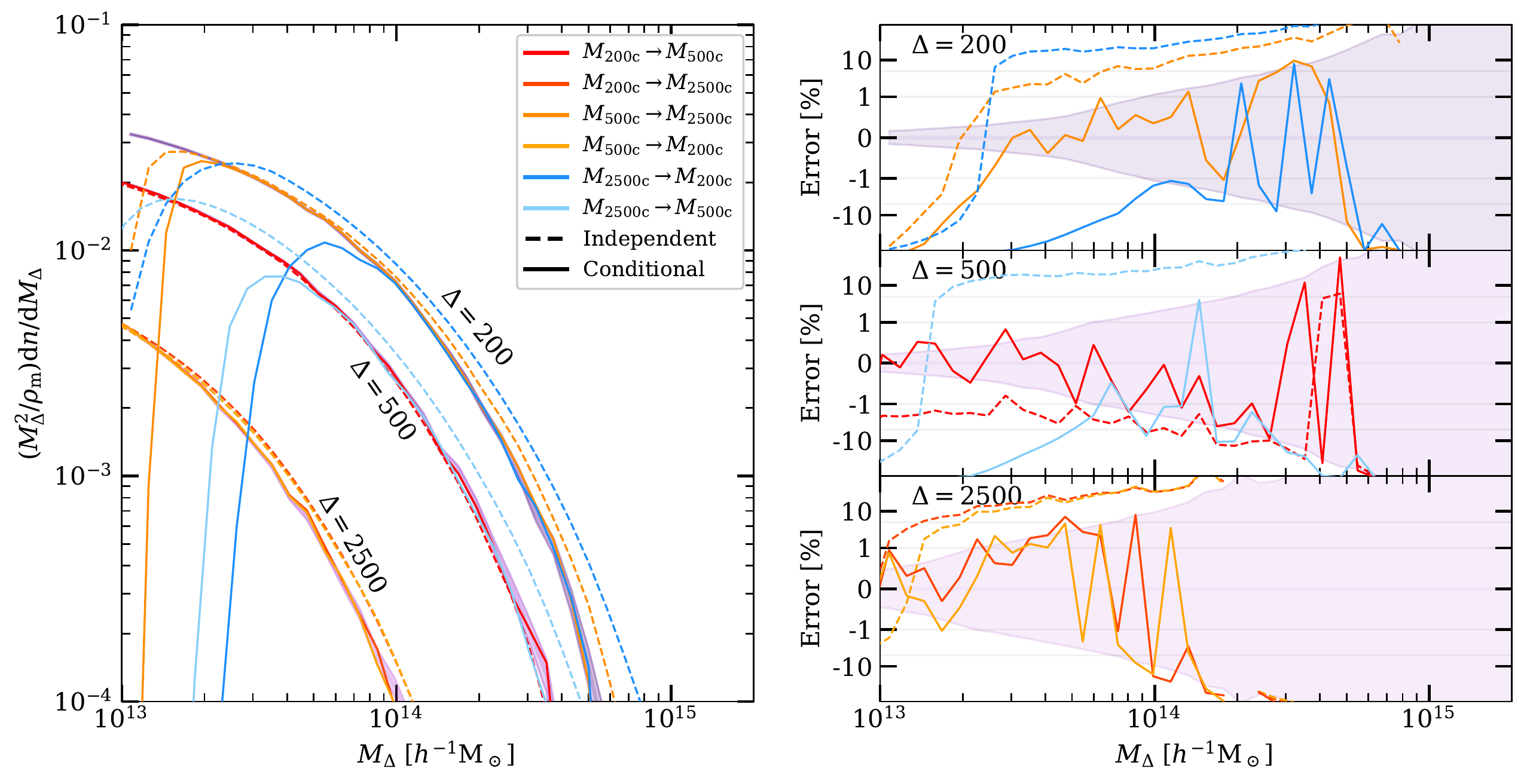}
    \caption{Same as Fig.~\ref{fig:reconstruction}, but at redshift $z= 1$. 
    }
    \label{fig:reconstruction_z1p03}
\end{figure*}

In Section~\ref{HMF_tests} we   test the validity of the inward and outward HMF reconstructions using the halo Uchuu catalogue at $z=0$. Here we present the results of  similar analyses for the halo catalogues at $z=0.5$ and $1$. These are summarised in the plots shown in Fig.~\ref{fig:reconstruction_z0p49} and Fig.~\ref{fig:reconstruction_z1p03}, respectively. We find the same trends as shown in Fig.~\ref{fig:reconstruction}. In particular, we note again that the use of the conditional sparsity distribution results in reconstructed HMFs that are within the statistical errors of those estimated from the N-body catalogues. This is not the case of the inward and outward reconstructions obtained using the sparsity marginal distribution.

\section{Profiles with more than one parameter}\label{app:Einastoprofile}

Within this work we   present an in-depth exploration of the relation between the distributions of NFW concentrations and sparsities. This methodology can be extended to profiles with more than one parameter describing the shape. Here we take the example of another widely used profile, the Einasto profile \citep{Einasto1965}
\begin{equation}
    \rho(r) = \rho_{-2}\exp\left\{-\frac{2}{\alpha}\left[\left(\frac{r}{r_{-2}}\right)^{-\alpha} - 1\right]\right\},
    \label{eq:Einasto}
\end{equation}
which has gained significant traction over the last decade. This profile is able to fit the density profiles of dark matter haloes to a greater acuracy than the NFW profile, even accounting for the fact that it has an additional parameter. However, using the Einasto profile    comes with the added complexity that  the mass profile can only be expressed numerically and not analytically.

Here the additional parameter increases the complexity of the transformation between the two parameters describing the shape of the profile, ($r_{-2}$, $\alpha$), and sparsity. For each pair ($r_{-2}$, $\alpha$) we fix $\rho_{-2}$ by fixing $M_{200 \rm c}$. Taking into account this constraint, we calculate the sparsity by solving
\begin{equation}
    r_{\Delta}^3 = \frac{3}{\Delta \rho_{\rm c}}\int^{r_\Delta}_0 r^2\rho(r; \rho_{-2}, r_{-2},\alpha)\dd r
\end{equation}
for both values of $\Delta$. This results in a mass dependent transformation between the Einasto parameters and $s_{\Delta_1, \Delta_2}$.

To transform the distribution of Einasto profile parameters into a distribution of sparsities we choose, in the conventions of Appendix~\ref{app:random_variables}, $Z = s_{\Delta_1, \Delta_2}$, $X = r_{-2}$, and $W = Y = \alpha$, which considerably simplifies the expression of the Jacobian,
\begin{equation}
    \rho_{s, \alpha}(s,\alpha) = \rho_{r_{-2},\alpha}[g_{r_{-2}}(s,\alpha), \alpha]\left|\partial_s g_{r_{-2}}\right|,
\end{equation}
where, as for the mass profile, the expression $g_{r_{-2}}(s,\alpha)$ has to be estimated numerically. This function can simply be seen as the value of $r_{-2}$ for a given value of $s_{\Delta_1,\Delta_2}$ and $\alpha$. The PDF of sparsity is then
\begin{equation}
    \rho_{\rm s}(s) = \int \rho_{r_{-2},\alpha}[g_{r_{-2}}(s,\alpha), \alpha]\left|\partial_s g_{r_{-2}}\right|\dd \alpha.
\end{equation}
This methodology can  be extended to any number of parameters, however   with  the complexity of having $n-1$ dimensional integrals for a profile with $n$ parameters.
\end{appendix}

\end{document}